\documentclass[aps,superscriptaddress,eqsecnum,nofootinbib,preprintnumbers]{revtex4}
\usepackage{graphicx,epsfig}
\usepackage{amsmath}
\usepackage {amssymb}
\usepackage{subfigure}

\usepackage{relsize}

\newcommand{\be}{\begin{equation}}
\newcommand{\ee}{\end{equation}}
\newcommand{\bea}{\begin{eqnarray}}
\newcommand{\eea}{\end{eqnarray}}
\newcommand{\nn}{\nonumber}

\begin{document}

\title{Hyperscaling violating black holes in scalar-torsion theories
}

\author{Georgios Kofinas}
\email{gkofinas@aegean.gr} \affiliation{Research Group of Geometry,
Dynamical Systems and Cosmology\\
Department of Information and Communication Systems Engineering\\
University of the Aegean, Karlovassi 83200, Samos, Greece}

\begin{abstract}
We study a gravity theory where a scalar field with potential, beyond its minimal coupling, is also
coupled through a non-minimal derivative coupling with the torsion scalar which is the teleparallel
equivalent of Einstein gravity. This theory provides second order equations of motion and we find
large-distance non-perturbative static spherically symmetric four-dimensional solutions. Among them a
general class of black hole solutions is found for some range of the parameters/integration constants
with asymptotics of the form of hyperscaling violating Lifshitz spacetime with spherical horizon
topology. Although the scalar field diverges at the horizon, its energy density and pressures are finite
there. From the astrophysical point of view, this solution provides extra deflection of light
compared to the Newtonian deflection.

\end{abstract}

\maketitle

\section{Introduction} \label{Introduction}

Teleparallelism \cite{ein28,TEGR,TEGR22,Hayashi:1979qx,JGPereira,Arcos:2005ec,Maluf:2013gaa,Pereira:2013qza}
is an equivalent approach of General Relativity with deep implications
which cannot be provided by the standard metric formulation of gravity. For example, it provides
a correct definition for the local energy-momentum tensor of the gravitational field which cannot
be expressed in terms of the metric. This formulation is based on the use of the vierbein, along
with a connection of vanishing curvature, as basic objects, not aiming to define another
gravitational theory, but to describe anew Einstein's gravity itself. Torsion turns out to be the
appropriate quantity for this reformulation. The Einstein-Hilbert Lagrangian is expressed (up
to boundary quantities) in terms of torsion, so that only up to first order derivatives in the dynamical
quantities (vierbein/connection) appear in the new Lagrangian. This Lagrangian is a particular
quadratic combination of torsion which is called torsion scalar $T$. As a result, the Einstein
equations have also their torsion analogue. As the metric formulation, also the teleparallel one is
both diffeomorphism and local Lorentz invariant. By employing the Weitzenb\"{o}ck connection
as solution of the connection equation of motion (vanishing curvature, but non-vanishing antisymmetric
piece of the connection when expressed in coordinates, and hence non-vanishing torsion), the
vierbein remains as the single dynamical variable and only the Lagrangian lacks local Lorentz invariance.
Notice that in practice the teleparallelism condition of vanishing curvature is implemented through the
Weitzenb\"{o}ck connection, and therefore the teleparallel approach is different than
other approaches (e.g. \cite{Mardones:1990qc}) where torsion introduces a new dynamical field beyond
the fields (metric, vierbein) describing the Einstein sector. Not only Einstein gravity but also
Gauss-Bonnet gravity has been shown to possess its teleparallel representation through another torsion
scalar $T_{G}$ \cite{Kofinas:2014owa}. Modifications of Einstein or Gauss-Bonnet gravity, different than
the curvature based modifications $f(R)$ \cite{DeFelice:2010aj,Nojiri:2010wj},
$f(R,G)$ \cite{Nojiri:2005jg,DeFelice:2008wz,Davis:2007ta} have been constructed based on $T$, $T_{G}$,
i.e. $f(T)$ \cite{Ferraro:2006jd,Linder:2010py,Chen:2010va,Zhang:2011qp,Aviles:2013nga},
$f(T,T_{G})$ \cite{Kofinas:2014daa} theories have been investigated which in the covariant formulation
are also diffeomorphism and local Lorentz invariant.

Beyond the purely gravitational sector, there are in nature various sorts of matter fields. Scalar
fields for example are predicted by fundamental theories and may be present in the universe at large
scales mixed up with baryonic matter, and so be present in ordinary stars. Spherically symmetric
configurations for self-gravitating scalar fields with various types of interactions have been derived
as solutions of the coupled Einstein-scalar system \cite{BBMB,Martinez:2004nb,Martinez:2005di}. At
cosmological scales the scalar field dynamics allows to investigate the features of the early universe,
or a quintessence scalar field can be interpreted as the present dynamical dark energy sector
\cite{Sahni:1998at}. The appearance of a non-minimal coupling of the form $f(\phi)R$ is motivated by
many reasons, such as the variability of the fundamental constants, the Kaluza-Klein compactification
scheme, the low-energy limit of superstring theory, e.tc. A non-minimal derivative coupling between
matter and gravity is the next step of generalization, particularly if the Newton's constant depends
on the gravitational field source mass \cite{Amendola:1993uh}.

In \cite{Kofinas:2015hla}, a first step was made in an attempt to probe the effect of torsion to General
Relativity when this is treated in its teleparallel reformulation, so the torsion scalar $T$ was coupled
to a scalar field $\phi$ in four spacetime dimensions. Since for a diagonal vierbein,
a simpler coupling of the form $\phi T$ does not possess spherically symmetric solutions (however
for three dimensions see \cite{Gonzalez:2014pwa}), a non-minimal derivative coupling of the form
$Tg^{\mu\nu}\partial_{\mu}\phi\,\partial_{\nu}\phi$ was considered. This is a novel coupling which
provides only second order equations of motion. The metric-based analogous coupling of the form
$Rg^{\mu\nu}\partial_{\mu}\phi\,\partial_{\nu}\phi$ gives higher order equations of motion
\cite{Amendola:1993uh}, while the sector $G^{\mu\nu}\partial_{\mu}\phi\,\partial_{\nu}\phi$ of the
general Horndeski Lagrangian also is a healthy theory with second order field equations \cite{horny}.
Of course, the above derivative coupling between torsion and matter is not the only one that can be
constructed, but other combinations of $\partial_{\mu}\phi\,\partial_{\nu}\phi$ with quadratic (of even
higher) powers of torsion could be presented. In \cite{Kofinas:2015hla} static spherically symmetric
solutions were studied for the above coupling with torsion, however after the formulation was developed,
only a general class of large-distance linearized solutions around its asymptotic Anti-de Sitter (AdS)
form was found. Here, we will extend the analysis and find other (general and special) large-distance
solutions in the non-linearized regime however, which turn out to be black hole solutions of the theory.
The approximated solutions found have the meaning that they become of higher and higher accuracy
as we restrict in particular regions of the parameter/integration constants space, which is not
however fine-tuned or particularly narrow. The regime of applicability of these solutions is for
any distance larger than some length scale determined by these parameters.
A general such black hole solution turns out to have asymptotics
which deviate from the AdS form and are of the hypescaling violating
Lifshitz form.

Our work is organized as follows. In Section II the basic elements and equations of the model appeared
in \cite{Kofinas:2015hla} are briefly presented so that to make the present work autonomous. In Section
III we proceed with the essential steps in order to simplify the master equation of the problem and
facilitate our quest for finding large-distance non-perturbative spherically symmetric solutions of the
theory. In Section IV we consider a special solution (i.e. for a particular value of the integration
constant of the scalar field) and find black holes and other solutions along with the scalar field
profile and the potential. In Section V a dynamical systems analysis reveals the global phase portrait
of all the local solutions and serves to obtain an intuition for the approximations made in order to
obtain the solutions of the next Section. In Section VI a general from the viewpoint of the number
of integration constants black hole solution is found, its asymptotic behavior is analyzed and some
astrophysical perspective of the solution is elaborated. Finally, Section VII is devoted to our
conclusions.

\section{General formulation}
\label{einstein}

We will consider the following non-minimal derivative coupling of a scalar field $\phi$ with torsion
\cite{Kofinas:2015hla}
\begin{equation}
S=-\frac{1}{2\kappa^{2}}\int d^{4}\!x\,e\,T-\int d^{4}\!x\,e\,
\Big[\Big(\frac{1}{2}-\xi T\Big)g^{\mu\nu}
\partial_{\mu}\phi\,\partial_{\nu}\phi+V\Big]\,,
\label{Tnonminimal}
\end{equation}
where $V$ is the potential of the scalar field and the parameter $\xi$ has dimensions of length square,
so $\sqrt{|\xi|}$ introduces a new length scale. The torsion scalar $T$ is defined by
\begin{eqnarray}
T=\frac{1}{4}T^{\mu\nu\lambda}T_{\mu\nu\lambda}+\frac{1}{2}T^{\mu\nu\lambda}
T_{\lambda\nu\mu}-T_{\nu}^{\,\,\,\nu\mu}T^{\lambda}_{\,\,\,\lambda\mu}
=S^{\mu\nu\lambda}T_{\mu\nu\lambda}\,,
\label{Tquad}
\end{eqnarray}
where the tensor $S^{\mu\nu\lambda}$ is
\begin{equation}
S^{\mu\nu\lambda}=\frac{1}{2}\mathcal{K}^{\nu\lambda\mu}+\frac{1}{2}(g^{\mu\lambda}T_{\rho}^{\,\,\,\rho\nu}
-g^{\mu\nu}T_{\rho}^{\,\,\,\rho\lambda})=-S^{\mu\lambda\nu}
\label{fyt}
\end{equation}
and
\begin{equation}
\mathcal{K}_{\mu\nu\lambda}=\frac{1}{2}(T_{\lambda\mu\nu}-T_{\nu\lambda\mu}-T_{\mu\nu\lambda})
=-\mathcal{K}_{\nu\mu\lambda}=\omega_{\mu\nu\lambda}-\Gamma_{\mu\nu\lambda}
\label{lji}
\end{equation}
is the contortion tensor. The connection $\omega^a_{\,\,\,b}=\omega^a_{\,\,\,b\mu}dx^\mu=\omega^a_{\,\,\,bc}e^c$
is assumed to satisfy the teleparallelism condition of vanishing curvature
$R^{a}_{\,\,\, b\mu\nu}\!=\omega^{a}_{\,\,\,b\nu,\mu}-
\omega^{a}_{\,\,\,b\mu,\nu}+\omega^{a}_{\,\,\,c\mu}\omega^{c}_{\,\,\,b\nu}-\omega^{a}_{\,\,\,c\nu}
\omega^{c}_{\,\,\,b\mu}=0$ (indices $a,b,...$ refer to tangent space, while $\mu,\nu,...$ are
coordinate ones). In terms of an arbitrary orthonormal vierbein $e_a=e^{\,\,\, \mu}_a\partial_\mu$
(with dual $e^a=e^a_{\,\,\, \mu}d x^\mu$ and $e=\text{det}(e^{a}_{\,\,\,\mu})$),
i.e. $g_{\mu\nu}=\eta_{ab}\,e^a_{\,\,\,\mu}\,e^b_{\,\,\,\nu}$
where $\eta_{ab}=\text{diag}(-1,1,1,1)$ is the Minkowski metric, the torsion is defined by
$T^{a}_{\,\,\,bc}=\omega^{a}_{\,\,\,cb}-\omega^{a}_{\,\,\,bc}-e_{b}^{\,\,\,\mu}
e_{c}^{\,\,\,\nu}(e^{a}_{\,\,\,\mu,\nu}-e^{a}_{\,\,\,\nu,\mu})$ and is a tensor under local
Lorentz transformations and under diffeomorphisms. The metric is assumed to be compatible
with the connection $\omega^{a}_{\,\,\,bc}$, i.e. $\eta_{ab|c}=0$, and so $\omega_{abc}=-\omega_{bac}$,
where $|$ denotes covariant differentiation with respect to $\omega^{a}_{\,\,\,bc}$. The importance
of the torsion scalar $T$ is that it provides after variation with respect to $e_{a}^{\,\,\,\mu}$ the
Einstein equations, i.e. $eT$ equals $e\bar{R}$ up to a total divergence, where
$\bar{R}$ is the Ricci scalar of the Christoffel connection $\Gamma^{\mu}_{\,\,\,\nu\lambda}$.

A curvature-based analogue of the above non-minimal coupling is the term
$Rg^{\mu\nu}\partial_{\mu}\phi\,\partial_{\nu}\phi$. The basic difference between these two is that
while the curvature coupling provides higher than second derivatives (and therefore ghosts), here,
the torsion coupling gives only second order equations of motion.
In some sense, the coupling in (\ref{Tnonminimal}) can be said that it is closer to the curvature
coupling $G^{\mu\nu}\partial_{\mu}\phi\,\partial_{\nu}\phi$ which also provides second order
equations of motion, and indeed at the level only of background cosmology these two couplings
coincide.

The equations of motion for the vierbein and the scalar field were found in \cite{Kofinas:2015hla} to be
\begin{eqnarray}
&&\!\!\!\!\!\!\!\!\!\!\!\!\!\!\Big(\frac{2}{\kappa^{2}}\!-\!4\xi\phi_{,\rho}\phi^{,\rho}\Big)
\Big[(eS_{\kappa}^{\,\,\,\lambda\nu} e_{b}^{\,\,\,\kappa})_{,\nu}e^{b}_{\,\,\,\mu}
+e\Big(\frac{1}{4}T\delta^{\lambda}_{\mu}\!-\!S^{\nu\kappa\lambda}T_{\nu\kappa\mu}\Big)\Big]
+4\xi\Big[\frac{1}{2}eT\phi_{,\mu}\phi^{,\lambda}
\!+\!eS_{\mu}^{\,\,\,\nu\lambda}(\phi_{,\kappa}\phi^{,\kappa})_{,\nu}
\Big]\nn\\
&&\,\,\,\,\,\,\,\,\,\,\,\,\,\,\,\,\,\,\,\,\,\,\,\,\,\,\,\,\,\,\,\,\,
+e\Big(\frac{1}{2}\phi_{,\rho}\phi^{,\rho}\delta^{\lambda}_{\mu}-\phi_{,\mu}
\phi^{,\lambda}+V\delta^{\lambda}_{\mu}\Big)
-\Big(\frac{2}{\kappa^{2}}\!-\!4\xi\phi_{,\rho}\phi^{,\rho}\Big)
eS^{dca}\omega_{bdc}e_{a}^{\,\,\,\lambda}e^{b}_{\,\,\,\mu}=0
\label{equationeomega1}
\end{eqnarray}
\begin{equation}
\big[e(1-2\xi T)\phi^{,\mu}\big]_{,\mu}-e\frac{dV}{d\phi}=0\,,
\label{equationphi}
\end{equation}
where $\phi^{,\mu}=g^{\mu\nu}\phi_{,\nu}$. As for the connection $\omega_{\,\,\,bc}^{a}$, we adopt
the Weitzenb\"{o}ck solution of the equations of motion $R_{abcd}(\omega^{a}_{\,\,\,bc})=0$, i.e.
$\omega_{\,\,\,bc}^{a}=0$ in the frame $e_{a}^{\,\,\,\mu}$ we will work with, therefore,
$\omega_{\,\,\,\mu\nu}^{\lambda}=e_{a}^{\,\,\,\lambda}e^{a}_{\,\,\,\mu,\nu}$ in all coordinate frames.

We are interested in extracting static spherically symmetric solutions of the four-dimensional
scalar-torsion gravity described by the action (\ref{Tnonminimal}). Due to the spherical symmetry,
the general form of the metric is diagonal
\begin{align}
ds^{2}=-N(r)^2 dt^2+K(r)^{-2}dr^2 +R(r)^2 d\Omega_{2}^2\,,
\label{sphermetric}
\end{align}
where $d\Omega_{2}^2 =d\theta^2+\sin^2\!\theta\,d\varphi^2$ is the two-dimensional sphere, and
$N(r)$, $K(r)$ and $R(r)$ are three unknown functions. One could have chosen the gauge with $R(r)=r$,
but it turns out to be more convenient to let the function $R(r)$ free. As a next step in order
to extract a vierbein that gives rise to the above metric, we choose the simplest diagonal
vierbein of the form
\begin{equation}
e^a_{\,\,\,\mu} ={\text{diag}}\left(N(r),K(r)^{-1},R(r),R(r)\sin\!\theta\right).
\label{vierb1}
\end{equation}
Substitution of (\ref{vierb1}) in the field equations (\ref{equationeomega1}), (\ref{equationphi})
provides the following set of equations of motion
\begin{eqnarray}
\!\!\!\!\!\!
\frac{1}{K^{2}}\Big(\phi'^{2}\!+\!\frac{2V}{K^{2}}\Big)+2\Big(\frac{1}{\kappa^{2}K^{2}}
\!-\!2\xi\phi'^{2}\Big)
\Big(\frac{R'^{2}}{R^{2}}\!+\!\frac{2R''}{R}\!+\!2\frac{R'}{R}\frac{K'}{K}\!-\!
\frac{1}{K^{2}R^{2}}\Big)-16\xi\phi'\frac{R'}{R}\Big(\frac{K'}{K}\phi'\!+\!\phi''\Big)&\!\!=\!\!&0
\label{1}\\
\!\!\!\!\!\!
8\xi\phi'^{2}\frac{R'}{R}\Big(\frac{R'}{R}\!+\!\frac{2N'}{N}\Big)+\frac{1}{K^{2}}\Big(\phi'^{2}
\!-\!\frac{2V}{K^{2}}\Big)+2\Big(2\xi\phi'^{2}\!-\!\frac{1}{\kappa^{2}K^{2}}\Big)\Big[\frac{R'}{R}
\Big(\frac{R'}{R}\!+\!\frac{2N'}{N}\Big)\!-\!\frac{1}{K^{2}R^{2}}\Big]&\!\!=\!\!&0\label{2}\\
\!\!\!\!\!\!
\frac{1}{K^{2}}\Big(\phi'^{2}\!+\!\frac{2V}{K^{2}}\Big)+2\Big(\frac{1}{\kappa^{2}K^{2}}
\!-\!2\xi\phi'^{2}\Big)
\Big[\frac{N'}{N}\Big(\frac{R'}{R}\!+\!\frac{K'}{K}\Big)\!+\!\frac{R'}{R}\frac{K'}{K}\!+\!\frac{N''}{N}
\!+\!\frac{R''}{R}\Big]
-8\xi\phi'\Big(\frac{R'}{R}\!+\!\frac{N'}{N}\Big)\Big(\frac{K'}{K}\phi'\!+\!\phi''\Big)&\!\!=\!\!&0
\label{3}\\
\phi'\Big(\frac{K'}{K}\phi'\!+\!\phi''\Big)&\!\!=\!\!&0\label{4}\\
\Big{\{}KNR^{2}\phi'\Big[1+4\xi K^{2}\frac{R'}{R}\Big(\frac{R'}{R}\!+\!\frac{2N'}{N}\Big)\Big]\Big{\}}'
-\frac{NR^{2}}{K}\frac{dV}{d\phi}&\!\!=\!\!&0\label{eqmsc}\,,
\end{eqnarray}
where a prime denotes differentiation with respect to $r$. Equations (\ref{1})-(\ref{4}) are
respectively the non-vanishing $(\lambda,\mu)=tt,rr,\theta\theta,\theta r$ components of the system
(\ref{equationeomega1}). An interesting characteristic, usual in
teleparallel gravity, is the appearance of the off-diagonal equation (\ref{4}), although the metric
ansatz (\ref{sphermetric}) and the vielbein (\ref{vierb1}) are diagonal. This does not happen in the
equations of motion derived from curvature-based actions, and in general, the extra equation imposes difficulties in
finding solutions in the torsion formulation. Equations (\ref{1})-(\ref{eqmsc}) have the reparametrizion
invariance, so for $r\rightarrow \tilde{r}(r)$ and $K\rightarrow K\frac{d\tilde {r}}{dr}$,
$N\rightarrow N$, $R\rightarrow R$, $\phi\rightarrow \phi$, the equations remain form invariant. This
implies that one of these equations is not necessary because it arises from the others. Thus, we
remain with four independent equations for four unknowns $N(r), K(r), \phi(r), V(r)$
($R(r)$ is not considered in the enumeration of the unknowns due to the choice of the radial gauge),
so there is no arbitrary function left. As a result the previous system has basically a unique
solution and this gives us a hope to extract some information on the structure and
behaviour of the derived solutions not constrained by a prefixed potential.

In \cite{Kofinas:2015hla} it was shown that if
\begin{eqnarray}
&&x=\ln{R}\label{xsmall}\\
&&Y=\Big(\frac{\dot{R}}{R}\Big)^{2}\,,\label{zsmall}
\end{eqnarray}
where a dot denotes differentiation with respect to $\phi$, the system of equations
(\ref{1})-(\ref{eqmsc}) reduces to the master equation for $Y(x)$
\begin{equation}
2\frac{d^{2}Y}{dx^{2}}-\frac{1}{Y}\Big(\frac{dY}{dx}\Big)^{2}+2\Big(2-\frac{\eta\nu^{2}}{Y}\Big)\frac{dY}{dx}
+3Y+2\eta\nu^{2}-12\frac{\eta^{2}}{\tilde{\eta}^{2}}Y\frac{\frac{dY}{dx}+3Y-\frac{2}{3\nu^{2}}e^{-2x}}
{\frac{dY}{dx}+3Y+2\eta\nu^{2}}=0\,,
\label{jsi}
\end{equation}
where
\begin{equation}
\eta=\frac{\kappa^{2}}{2\nu^{2}(1-2\xi\kappa^{2}\nu^{2})}\,\,\,\,\,\,\,,\,\,\,\,\,\,\,
\tilde{\eta}=\frac{\kappa^{2}}{\nu^{2}(1-6\xi\kappa^{2}\nu^{2})}\,.
\label{hud}
\end{equation}
The integration constant $\nu$ with dimensions of inverse length square is introduced through the
relation $\phi'=\frac{\nu}{K}$ arising from (\ref{4}). After equation
(\ref{jsi}) is solved, the potential, the scalar field profile and the metric can be found.
Namely, the potential $V(R)$ is found from
\begin{equation}
V=\frac{1}{2\eta\nu^{2}}e^{-2x}-\frac{\nu^{2}}{2}-\frac{1}{2\eta}\Big(\frac{dY}{dx}+3Y\Big)
\label{wij}
\end{equation}
and the scalar field $\phi(R)$ is obtained from
\begin{equation}
\Big(\frac{dx}{d\phi}\Big)^{2}=Y(x)\,.
\label{dxdphiv}
\end{equation}
The lapse function $N(R)$ is found from
\begin{equation}
\Big[\frac{d\ln(RN^2)}{dx}\Big]^{2}=\frac{Z(x)}{Y(x)}\,,
\label{dlnRN21}
\end{equation}
where
\begin{equation}
Z=\frac{\tilde{\eta}^{2}}{4\eta^{2}Y}\Big(\frac{dY}{dx}+3Y+2\eta\nu^{2}\Big)^{2}\,.
\label{gij}
\end{equation}
Finally the metric takes the form
\begin{equation}
ds^{2}=-N^{2}dt^{2}+\frac{dR^{2}}{\nu^{2}R^{2}Y}+R^{2}d\Omega_{2}^{2}\,.
\label{jsb}
\end{equation}
Note that as far as $\xi$ and $\nu$ remain unrelated, the coupling $\xi$ appears only through the
combination $\xi\kappa^{2}\nu^{2}$ and this is due to the initial interaction term
$Tg^{\mu\nu}\partial_{\mu}\phi\partial_{\nu}\phi$.

It is also interesting to investigate the behavior of another physically meaningful quantity which is
the energy-momentum tensor $\mathcal{T}^{a}_{\,\,\,\,\mu}$ of the scalar field $\phi$. This arises
by variation with respect to $e_{a}^{\,\,\,\mu}$ of the scalar field action, i.e. of the second
integral appearing in (\ref{Tnonminimal}). One can find from (\ref{1})-(\ref{4}) the on-shell
components of $\mathcal{T}^{\nu}_{\,\,\,\,\mu}$ as
$\mathcal{T}^{t}_{\,\,\,\,t}=\mathcal{T}^{\theta}_{\,\,\,\,\theta}=
\frac{1}{2}\frac{2V+K^{2}\phi'^{2}}{2\xi\kappa^{2}K^{2}\phi'^{2}-1}$,
$\mathcal{T}^{R}_{\,\,\,\,R}=\mathcal{T}^{r}_{\,\,\,\,r}=\frac{1}{2}
\frac{2V-(1+8\xi R^{-2})K^{2}\phi'^{2}}{6\xi\kappa^{2}K^{2}\phi'^{2}-1}$,
which express the energy density and the pressures of the scalar field. Since $K\phi'=\nu$,
it arises $\mathcal{T}^{t}_{\,\,\,\,t}=\mathcal{T}^{\theta}_{\,\,\,\,\theta}=\frac{1}{2}\frac{2V+\nu^{2}}{2\xi\kappa^{2}
\nu^{2}-1}$, $\mathcal{T}^{R}_{\,\,\,\,R}=
\mathcal{T}^{r}_{\,\,\,\,r}=\frac{1}{2}\frac{2V-\nu^{2}(1+8\xi R^{-2})}{6\xi\kappa^{2}
\nu^{2}-1}$.

\section{Simplifications}

We start by writing equation (\ref{jsi}) as
\begin{equation}
2\frac{d^{2}Y}{dx^{2}}-\frac{1}{Y}\Big(\frac{dY}{dx}\Big)^{2}+2\Big(2-\frac{\eta\nu^{2}}{Y}\Big)
\frac{dY}{dx}+3\Big(1-\frac{4\eta^{2}}{\tilde{\eta}^{2}}\Big)Y+2\eta\nu^{2}
+24\frac{\eta^{2}}{\tilde{\eta}^{2}}Y\frac{\eta\nu^{2}+\frac{1}{3\nu^{2}}e^{-2x}}
{\frac{dY}{dx}+3Y+2\eta\nu^{2}}=0\,,
\label{akf}
\end{equation}
which also takes the form
\begin{equation}
4\sqrt{Y}\frac{d^{2}\sqrt{Y}}{dx^{2}}+2\Big(2-\frac{\eta\nu^{2}}{Y}\Big)\frac{dY}{dx}
+3\Big(1-\frac{4\eta^{2}}{\tilde{\eta}^{2}}\Big)Y+2\eta\nu^{2}
+24\frac{\eta^{2}}{\tilde{\eta}^{2}}Y\frac{\eta\nu^{2}+\frac{1}{3\nu^{2}}e^{-2x}}
{\frac{dY}{dx}+3Y+2\eta\nu^{2}}=0\,.
\label{kfi}
\end{equation}
If we define
\begin{equation}
U=\sqrt{Y}\,,
\label{kss}
\end{equation}
equation (\ref{kfi}) takes the form
\begin{equation}
\frac{d^{2}U}{dx^{2}}+\Big(2-\frac{\eta\nu^{2}}{U^{2}}\Big)\frac{dU}{dx}
+\frac{3}{4}\Big(1-\frac{4\eta^{2}}{\tilde{\eta}^{2}}\Big)U+\frac{\eta\nu^{2}}{2U}
+\frac{3\eta^{2}}{\tilde{\eta}^{2}}\frac{\eta\nu^{2}+\frac{1}{3\nu^{2}}e^{-2x}}
{\frac{dU}{dx}+\frac{3}{2}U+\frac{\eta\nu^{2}}{U}}=0\,,
\label{dhj}
\end{equation}
which is simpler than the initial equation (\ref{jsi}).

Equation (\ref{dhj}) for
\begin{equation}
R\gg\frac{1}{\nu^{2}\sqrt{|\eta|}}
\label{ela}
\end{equation}
becomes autonomous
\begin{equation}
\frac{d^{2}U}{dx^{2}}+\Big(2-\frac{\eta\nu^{2}}{U^{2}}\Big)\frac{dU}{dx}
+\frac{3}{4}\Big(1-\frac{4\eta^{2}}{\tilde{\eta}^{2}}\Big)U+\frac{\eta\nu^{2}}{2U}
+\frac{3\eta^{2}}{\tilde{\eta}^{2}}\frac{\eta\nu^{2}}
{\frac{dU}{dx}+\frac{3}{2}U+\frac{\eta\nu^{2}}{U}}=0\,.
\label{ujf}
\end{equation}
If
\begin{equation}
\Omega=\frac{dU}{dx}+\frac{3}{2}U+\frac{\eta\nu^{2}}{U}\,,
\label{hdh}
\end{equation}
equation (\ref{ujf}) becomes
\begin{equation}
\Omega\Big(\Omega-\frac{3}{2}U-\frac{\eta\nu^{2}}{U}\Big)\frac{d\Omega}{dU}
+\frac{1}{2}\Omega^{2}-\frac{3\eta^{2}}{\tilde{\eta}^{2}}U\Omega+\frac{3\eta^{2}}{\tilde{\eta}^{2}}
\eta\nu^{2}=0\,.
\label{ujd}
\end{equation}
It is convenient to define $\alpha=\eta\nu^{2}$, $\beta=\frac{3\eta^{2}}{\tilde{\eta}^{2}}>0$. The
parameters $\alpha,\beta$ are not independent, but they obey the relations
$\alpha=\frac{\kappa^{2}}{2(1-2\Xi)}$,
$\beta=\frac{3}{4}\big(\frac{1-6\Xi}{1-2\Xi}\big)^{2}=
\frac{3}{4}\big(\frac{4\alpha}{\kappa^{2}}\!-\!3\big)^{2}$,
where
$\Xi=\xi\kappa^{2}\nu^{2}$. It is seen that for $\alpha>0$, the parameter $\beta$ can take any positive
value, while for $\alpha<0$ it is $\beta>\frac{27}{4}$. The inequality (\ref{ela}) gets the form
\begin{equation}
R\gg \frac{1}{|\nu|\sqrt{|\alpha|}}=\sqrt{2\Big|2\xi\!-\!\frac{1}{\kappa^{2}\nu^{2}}\Big|}\,.
\label{era}
\end{equation}
Finally, equation
(\ref{ujd}) becomes
\begin{equation}
\Omega\Big(\Omega-\frac{3}{2}U-\frac{\alpha}{U}\Big)\frac{d\Omega}{dU}
+\frac{1}{2}\Omega^{2}-\beta U\Omega+\alpha\beta=0\,.
\label{ujm}
\end{equation}
Equation (\ref{ujm}) is also written as
\begin{equation}
\Big(\Omega-\frac{3}{2}U-\frac{\alpha}{U}\Big)\frac{d\Omega}{dU}
+\frac{1}{2}\Omega-\beta U+\frac{\alpha\beta}{\Omega}=0\,,
\label{shf}
\end{equation}
or also
\begin{equation}
\frac{d\Omega}{dU}=\frac{\beta U-\frac{1}{2}\Omega-\frac{\alpha\beta}{\Omega}}
{\Omega-\frac{3}{2}U-\frac{\alpha}{U}}\,.
\label{wdh}
\end{equation}
Although (\ref{wdh}) is a first order differential equation, significantly and unexpectedly simpler
than the initial equation (\ref{jsi}), there is no known method how to solve it. However, its
characteristic form with the presence of the inverse powers of $U,\Omega$ will
enable us to find ``large''-distance non-perturbative solutions. As always, the word large is meant
in comparison with scales and integration constants of the problem, and depending on them, the
corresponding distances can be relevant for a physical situation.

Note that the system of equations (\ref{1})-(\ref{eqmsc}) for vanishing scalar field $\phi=0$ and a
cosmological constant as the potential $V$ has the standard Schwarzschild-(A)dS solution. A vanishing
scalar field means that $\nu=0$ and then it is $\alpha=\frac{\kappa^{2}}{2}$,
$\beta=\frac{3}{4}$, $\frac{\eta}{\tilde{\eta}}=\frac{1}{2}$, while $Y,U$ of (\ref{kfi}), (\ref{dhj})
become infinite. Moreover, the autonomous equation (\ref{ujf}) cannot recover this limit since the
inequality (\ref{era}) is never satisfied. The reason is that the exponential factor in (\ref{dhj})
becomes significant in this limit. However, although the general solutions to be derived
will not be continuous deformation of Schwarzschild, this does not necessarily mean that they will not
capture some interesting phenomenology, for example they could be related to galactic scales or beyond.

\section{Special solutions with $\nu\rightarrow 0$}

In the limit $\nu\rightarrow 0$ the theory posseses solutions, including black holes, with non trivial
scalar field and potential. For $\nu\rightarrow 0$, the scalar field $\phi\rightarrow 0$ and
(\ref{dxdphiv}) is not defined due to that $d\phi$ vanishes. However, the ``normalized'' field
$\tilde{\phi}=\frac{\phi}{\nu}$ obeys $\tilde{\phi}'=K^{-1}$ which is a meaningful equation. Similarly,
although $U\rightarrow \infty$ in (\ref{dhj}), however this equation still makes sense. Indeed, if we
define the ``normalized'' variable $\mathcal{U}=\nu U$, (\ref{dhj}) becomes in this limit
\begin{equation}
\Big(\frac{d\mathcal{U}}{dx}+\frac{3}{2}\mathcal{U}\Big)\Big(\frac{d^{2}\mathcal{U}}{dx^{2}}
+2\frac{d\mathcal{U}}{dx}\Big)+\frac{1}{4}e^{-2x}=0\,,
\label{nkb}
\end{equation}
where there is no free parameter left in (\ref{nkb}). Then, (\ref{dxdphiv}) becomes
\begin{equation}
\frac{dx}{d\tilde{\phi}}=\pm\,\mathcal{U}
\label{rll}
\end{equation}
and equations (\ref{dlnRN21}), (\ref{gij}) are meaningful. Finally, the metric (\ref{jsb}) becomes
\begin{equation}
ds^{2}=-N^{2}dt^{2}+\frac{dR^{2}}{R^{2}\,\mathcal{U}^{2}}+R^{2}d\Omega_{2}^{2}\,.
\label{dkd}
\end{equation}
The potential $V$ is found from (\ref{wij}) to be
\begin{equation}
V=\frac{1}{\kappa^{2}}\Big(e^{-2x}-3\,\mathcal{U}^{2}-2\,\mathcal{U}\frac{d\mathcal{U}}{dx}\Big)\,.
\label{ery}
\end{equation}
We will find as a warm up large-distance solutions in the above limit.

For $F=e^{x}\mathcal{U}$ we get from (\ref{nkb})
\begin{equation}
\Big(\frac{dF}{dx}+\frac{1}{2}F\Big)\Big(\frac{d^{2}F}{dx^{2}}-F\Big)+\frac{1}{4}=0\,.
\label{ejk}
\end{equation}
If $v(F)=\frac{dF}{dx}$, then
\begin{equation}
\Big(v+\frac{1}{2}F\Big)\Big(v\frac{dv}{dF}-F\Big)+\frac{1}{4}=0\,,
\label{nkw}
\end{equation}
or equivalently
\begin{equation}
v\frac{dv}{dF}=F-\frac{1}{2F\!+\!4v}\,.
\label{wqk}
\end{equation}
The variables $F,v$ are dimensionless and we can check by numerically solving (\ref{wqk}) that
every solution extends to large values of $|F|,|v|$. Thus, for $|F|,|v|\gg 1$ we get from (\ref{wqk})
\begin{equation}
v\frac{dv}{dF}=F\,,
\label{ewl}
\end{equation}
with general solution
\begin{equation}
v^{2}=F^{2}-\sigma\,,
\label{vle}
\end{equation}
where $\sigma$ is integration constant. Then,
\begin{equation}
F=\frac{\sigma R^{2}\!+\!\tau^{2}}{2\tau R}\,\,\,\,\,\text{or}\,\,\,\,\,
F=\frac{\tau^{2}R^{2}\!+\!\sigma}{2\tau R}\,,
\label{elc}
\end{equation}
where $\tau$ is another integration constant. Depending on $\sigma,\tau$, this solution can indeed
correspond to large values of $R$, since we see that large values of $R$ can correspond to large values
of $|F|,|v|$. Integrating equation (\ref{dlnRN21}) we find $N$. Finally, rescaling the time $t$, there
arise two sorts of metrics
\begin{equation}
ds^{2}=-F^{2}dt^{2}+\frac{dR^{2}}{F^{2}}+R^{2}d\Omega_{2}^{2}\,,
\label{qwf}
\end{equation}
or
\begin{equation}
ds^{2}=-\frac{1}{R^{2}F^{2}}dt^{2}+\frac{dR^{2}}{F^{2}}+R^{2}d\Omega_{2}^{2}\,.
\label{aef}
\end{equation}
Thus, there exist four different metrics.

Depending on the constants $\sigma,\tau$, the metric (\ref{qwf}) may possess horizon.
To be concrete, let us consider the first case of (\ref{elc})
with $\tau>0,\sigma<0$ and $|\sigma|\ll 1$. Then, it is obvious that large values of $R$
correspond to $F\ll 1$ (or $|F|\gg 1$), which means $R\gg\frac{\tau}{2}$. From (\ref{vle}) it is
also $|v|\gg 1$, while the quantity $|2F\!+\!4v|$ is large even for $Fv<0$. On the other hand, the
horizon of (\ref{qwf}) corresponds to $F=0$,
which means $R_{\text{hor}}=\frac{\tau}{\sqrt{|\sigma|}}$. This value can be made arbitrarily larger
than $\frac{\tau}{2}$ by choosing $|\sigma|$ sufficiently small, and so the existence of the horizon
within our approximation has been shown. In this case, (\ref{qwf}) with the first of (\ref{elc})
is the black hole
\begin{equation}
ds^{2}=-\frac{\big(|\sigma|R^{2}\!-\!\tau^{2}\big)^{2}}{4\tau^{2}R^{2}}dt^{2}+
\frac{4\tau^{2}R^{2}}{\big(|\sigma|R^{2}\!-\!\tau^{2}\big)^{2}}dR^{2}+R^{2}d\Omega_{2}^{2}
\label{eef}
\end{equation}
and the solution is certainly valid outside the horizon. Asymptotically, it gets an AdS form
\begin{equation}
ds_{\infty}^{2}=-\frac{\sigma^{2}}{4\tau^{2}}R^{2}dt^{2}+\frac{4\tau^{2}}{\sigma^{2}}\frac{dR^{2}}
{R^{2}}+R^{2}d\Omega_{2}^{2}\,.
\label{hru}
\end{equation}
The length scale introduced here is $\ell_{\text{eff}}^{-2}=\frac{\sigma^{2}}{4\tau^{2}}$ and the
corresponding effective cosmological constant $\Lambda_{\text{eff}}=-3\ell_{\text{eff}}^{-2}$.
\newline
Similarly, for the second case of (\ref{elc}) with $\tau<0,\sigma<0$ and $|\sigma|\ll 1$, it is
obvious that large values of $R$ correspond to $F\ll 1$ (or $|F|\gg 1$), which means
$R\gg\frac{|\sigma|}{2|\tau|}$. From (\ref{vle}) it is also $|v|\gg 1$, while the quantity $|2F\!+\!4v|$
is large even for $Fv<0$. On the other hand, the horizon of (\ref{qwf}) corresponds to $F=0$, which means
$R_{\text{hor}}=\frac{\sqrt{|\sigma|}}{|\tau|}$. This value again can be made arbitrarily larger than
$\frac{|\sigma|}{2|\tau|}$ for $|\sigma|$ sufficiently small, and so in this case (\ref{qwf}) with
the second of (\ref{elc}) forms another black hole
\begin{equation}
ds^{2}=-\frac{\big(\tau^{2}R^{2}\!-\!|\sigma|\big)^{2}}{4\tau^{2}R^{2}}dt^{2}+
\frac{4\tau^{2}R^{2}}{\big(\tau^{2}R^{2}\!-\!|\sigma|\big)^{2}}dR^{2}+R^{2}d\Omega_{2}^{2}\,.
\label{eof}
\end{equation}
Asymptotically, it also gets another AdS form
\begin{equation}
ds_{\infty}^{2}=-\frac{\tau^{2}}{4}R^{2}dt^{2}+\frac{4}{\tau^{2}}\frac{dR^{2}}
{R^{2}}+R^{2}d\Omega_{2}^{2}
\label{hrv}
\end{equation}
with $\ell_{\text{eff}}^{-2}=\frac{\tau^{2}}{4}$.

Concerning the scalar field $\tilde{\phi}$, we find from (\ref{rll}) for the two black hole solutions
(\ref{eef}), (\ref{eof}) respectively
\begin{equation}
\tilde{\phi}-\tilde{\phi}_{0}=\pm\frac{\tau}{\sigma}\ln{\big(|\sigma| R^{2}\!-\!\tau^{2}\big)}
\,\,\,\,\,\text{or}\,\,\,\,\,
\tilde{\phi}-\tilde{\phi}_{0}=\pm\frac{1}{\tau}\ln{(\tau^{2}R^{2}\!-\!|\sigma|)}\,.
\label{elv}
\end{equation}
Notice that the scalar field diverges at the position of the horizon and also asymptotically.
The scalar field is a secondary hair since it does not introduce a new non-trivial integration constant
(i.e. $\phi$ is not varied independently of the black hole parameters).

Finally from (\ref{ery}), since $\frac{d\mathcal{U}}{dx}=\frac{v-F}{R}$, we find the potential $V$
respectively
\begin{equation}
V=\frac{\sigma^{2}}{\kappa^{2}\tau^{2}}\frac{F^{-2}\!-\!1\!+\!2\epsilon\sqrt{1\!+\!|\sigma| F^{-2}}}
{\big(1\!+\!\sqrt{1\!+\!|\sigma| F^{-2}}\big)^{2}}\,\,\,\,\,\text{or}\,\,\,\,\,
V=\frac{\tau^{2}}{\kappa^{2}}\frac{F^{-2}\!-\!1\!+\!2\epsilon\sqrt{1\!+\!|\sigma| F^{-2}}}
{\big(1\!+\!\sqrt{1\!+\!|\sigma| F^{-2}}\big)^{2}}
\label{cdf}
\end{equation}
where $\epsilon$ is a $\pm$ sign. From (\ref{elv}), converting $R$ to $\tilde{\phi}$, we can easily
find $F(\tilde{\phi})$ respectively
\begin{equation}
F^{-2}=\frac{4\tau^{2}}{|\sigma|}\Big[
e^{\mp\frac{\sigma}{\tau}(\tilde{\phi}-\tilde{\phi}_{0})}\!+\!
\tau^{2}e^{\mp \frac{2\sigma}{\tau}(\tilde{\phi}-\tilde{\phi}_{0})}\Big]\,\,\,\,\,
\text{or}\,\,\,\,\,
F^{-2}=4\tau^{2}\Big[
e^{\mp\tau(\tilde{\phi}-\tilde{\phi}_{0})}\!-\!
\sigma e^{\mp 2\tau(\tilde{\phi}-\tilde{\phi}_{0})}\Big].
\label{qjs}
\end{equation}
Substituting (\ref{qjs}) into (\ref{cdf}) we find $V(\tilde{\phi})$. Note that for any solution
with specific $\sigma,\tau$ the corresponding potential $V$ that supports the solution also depends on
$\sigma,\tau$. We could rescale $\tilde{\phi}$ to $\bar{\phi}=\frac{\sigma}{\tau}(\tilde{\phi}-
\tilde{\phi}_{0})$ or $\bar{\phi}=\tau(\tilde{\phi}-\tilde{\phi}_{0})$, but still $\sigma,\tau$
remain in $V$. At spatial infinity it is
$F^{-2}\rightarrow 0$, therefore the potential becomes for the two cases
$V\rightarrow\frac{(2\epsilon-1)\sigma^{2}}{4\kappa^{2}\tau^{2}}$ or
$V\rightarrow\frac{(2\epsilon-1)\tau^{2}}{4\kappa^{2}}$, and for $\epsilon=-1$ these values coincide
with the corresponding values of $\Lambda_{\text{eff}}/\kappa^{2}$. For $\epsilon=1$, these values
of $V$ are positive, therefore the cosmological constant coming from the potential in the action is
positive, while the effective cosmological constant $\Lambda_{\text{eff}}$ of asymptotically AdS space
is negative. This discrepancy is due to the growth of a non trivial profile of the scalar field which is
coupled to the torsion and modifies the asymptotic form of the spacetime. In \cite{Kofinas:2015hla} a
similar discrepancy between $\Lambda_{\text{eff}}$ and $V$ has been found, but not for positive $V$
which is the case here. A positive value of $V$ has the merit that it can be related to the vacuum
energy of spacetime. On the other hand, in \cite{Martinez:2004nb} and other solutions the
asymptotic value of $V$ is the same as $\Lambda_{\text{eff}}/\kappa^{2}$. Note also that at the horizon
the potential $V$ is finite. Concerning the components of the energy-momentum tensor we have
$\mathcal{T}^{t}_{\,\,\,\,t}=\mathcal{T}^{R}_{\,\,\,\,R}=\mathcal{T}^{r}_{\,\,\,\,r}=
\mathcal{T}^{\theta}_{\,\,\,\,\theta}=-V$. Therefore, although the scalar field diverges at the horizon,
its energy density and the pressures are finite there. Asymptotically, still the energy density of
the scalar field and the pressures remain finite. Finally, notice that due to the higher
order pole at the horizon, the temperature vanishes in analogy to the Reissner-Nordstrom solution.

\begin{figure}[!]
\includegraphics*[width=260pt, height=180pt]{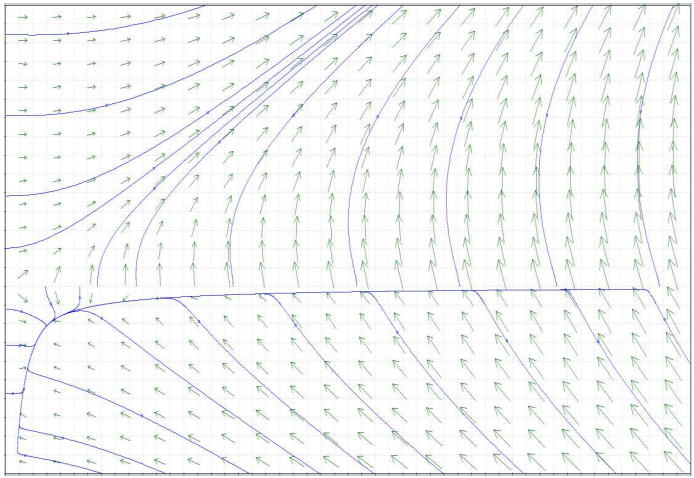}
\caption{
Phase portrait $(\hat{U},\hat{\Omega})$ for $\hat{\alpha}=-\frac{3}{4}$, $\beta=27$} \label{Fig. 1}
\end{figure}

\section{Dynamical systems analysis}

The second order autonomous differential equation (\ref{ujf}) can be converted into a two-dimensional
dynamical system with one function the variable $U$ and as second function some combination containing
the derivative $\frac{dU}{dx}$, e.g $\Omega$. To study the system it is better to convert into
dimensionless variables. We define
\begin{equation}
\hat{U}=\frac{U}{\kappa}\,\,\,\,,\,\,\,\,\hat{\Omega}=\frac{\Omega}{\kappa}\,\,\,\,,\,\,\,\,
\hat{\alpha}=\frac{\alpha}{\kappa^{2}}\,,
\label{eik}
\end{equation}
where $\hat{U}>0$ and $\hat{\Omega}$ is real.
So, instead of the equation (\ref{ujf}) we have the following equivalent system
\begin{eqnarray}
&&\frac{d\hat{U}}{dx}=\hat{\Omega}-\frac{3}{2}\hat{U}-\frac{\hat{\alpha}}{\hat{U}}
\label{qfh}\\
&&\frac{d\hat{\Omega}}{dx}=\beta \hat{U}-\frac{1}{2}\hat{\Omega}-
\frac{\hat{\alpha}\beta}{\hat{\Omega}}\,,
\label{dgh}
\end{eqnarray}
where $\beta=\frac{3}{4}(4\hat{\alpha}-3)^{2}$.
We can find the fixed points of this system by setting $\frac{d\hat{U}}{dx}=\frac{d\hat{\Omega}}{dx}=0$.
Then, we find these fixed points to be
\begin{equation}
\hat{U}_{\ast}=\sqrt{\frac{2\hat{\alpha}}{\pm 2\sqrt{3\beta}-3}}\,\,\,\,,\,\,\,\,\hat{\Omega}_{\ast}=\pm
\sqrt{\frac{6\hat{\alpha}\beta}{\pm 2\sqrt{3\beta}-3}}\,.
\label{hek}
\end{equation}
For $\hat{\alpha}<0$, $\beta>\frac{27}{4}$ there is exactly one fixed point
($\hat{U}_{\ast}=\sqrt{\frac{2|\hat{\alpha}|}{2\sqrt{3\beta}+3}}\,,\,\hat{\Omega}_{\ast}=-
\sqrt{\frac{6|\hat{\alpha}|\beta}{2\sqrt{3\beta}+3}}$). There are two negative eigenvalues
$3\lambda_{1,2}=-2(\sqrt{3\beta}+3)\pm\sqrt{12\beta+6\sqrt{3\beta}+9}$ of the linearized system,
thus the fixed point is attractor (stable node), as also seen in Fig. 1. In Fig. 1, 2, 3 the horizontal
axis is $\hat{U}$ and the vertical $\hat{\Omega}$, and the arrows show increasement of the radius $R$.
Moreover, due to the pole $\hat{\Omega}=0$ in (\ref{dgh}), the phase portraits are separated into
two quadrants, the upper one with $\hat{\Omega}>0$ and the lower with $\hat{\Omega}<0$.
For $\hat{\alpha}\in(0,\frac{1}{2})\cup (1,+\infty)$,
$\beta>\frac{3}{4}$ there is again exactly one fixed point
($\hat{U}_{\ast}=\sqrt{\frac{2\hat{\alpha}}{2\sqrt{3\beta}-3}}\,,\,\hat{\Omega}_{\ast}=
\sqrt{\frac{6\hat{\alpha}\beta}{2\sqrt{3\beta}-3}}$). In this case, the two eigenvalues
$3\lambda_{1,2}=2(\sqrt{3\beta}-3)\pm\sqrt{12\beta-6\sqrt{3\beta}+9}$ have opposite signs,
therefore the fixed point is saddle, as also seen in
Fig. 2. Finally, for $\frac{1}{2}<\alpha<1$, $0<\beta<\frac{3}{4}$ there is no fixed point, as
it is seen in Fig. 3. The values (\ref{hek}) of $\hat{U}$ are the same found in \cite{Kofinas:2015hla}
studying the asymptotic behaviour of equation (\ref{jsi}). From the above analysis it becomes clear
that only for $\hat{\alpha}<0$ (case A of \cite{Kofinas:2015hla}) the fixed point is attractor and
indeed corresponds to a large-distance asymptotic solution. For $\hat{\alpha}>0$ (case B of
\cite{Kofinas:2015hla}) the fixed point is saddle and does not correspond to an asymptotic solution.
It would be interesting for $\hat{\alpha}<0$ to investigate beyond the linearized asymptotic behavior
found in \cite{Kofinas:2015hla}, also the non-perturbative regime of the solution,
however, this will not be studied in the present paper. In the present paper we will find the
large-distance non-perturbative solutions which correspond to large values of $\hat{U},\hat{\Omega}$
as shown in the upper quadrants of Fig. 1 and 2. These solutions will be now shown that are attracted
by stable fixed points at infinity.

\begin{figure}[!]
\includegraphics*[width=260pt, height=180pt]{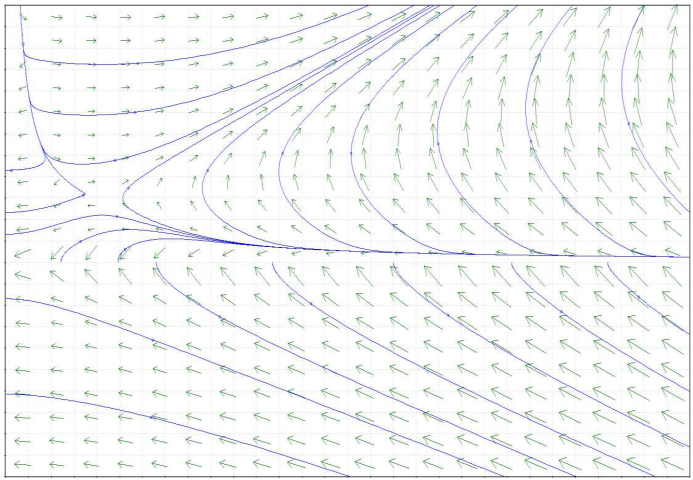}
\caption{
Phase portrait $(\hat{U},\hat{\Omega})$ for $\hat{\alpha}=\frac{3}{2}$, $\beta=\frac{27}{4}$} \label{Fig. 2}
\end{figure}

\begin{figure}[!]
\includegraphics*[width=260pt, height=180pt]{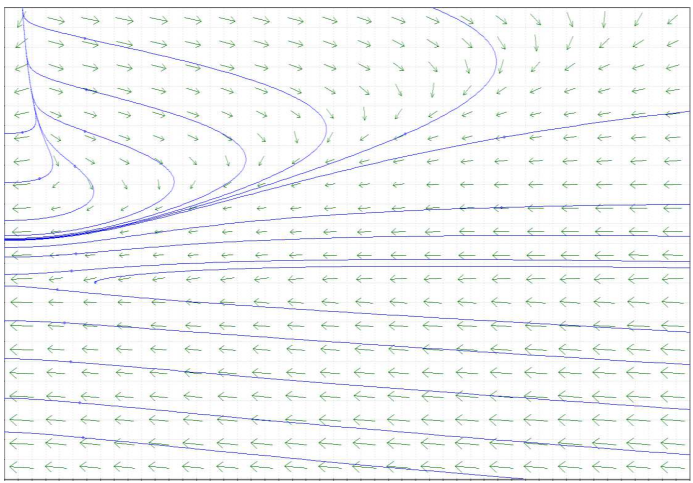}
\caption{
Phase portrait $(\hat{U},\hat{\Omega})$ for $\hat{\alpha}=\frac{7}{8}$, $\beta=\frac{3}{16}$} \label{Fig. 3}
\end{figure}

Due to that the dynamical system (\ref{qfh}), (\ref{dgh}) is non-compact, there could be non-trivial
fixed points at infinity, i.e. at the asymptotic region of the phase portrait $(\hat{U},\hat{\Omega})$.
These can be studied using the Poincar\'{e} projection method. We define the new variables (coordinates
of phase space) by
\begin{equation}
\hat{U}=\frac{\hat{r}}{1\!-\!\hat{r}}\cos{\hat{\theta}}\,\,\,\,\,\,,\,\,\,\,\,\,
\hat{\Omega}=\frac{\hat{r}}{1\!-\!\hat{r}}\sin{\hat{\theta}}
\label{qwfa}
\end{equation}
with $-\frac{\pi}{2}\leq\hat{\theta}\leq\frac{\pi}{2}$, $0\leq \hat{r}<1$.
The upper quadrants of Fig. 1, 2, 3 correspond to $\hat{\theta}>0$, while the lower ones to
$\hat{\theta}<0$. The limit $\hat{r}
\rightarrow 1^{-}$ corresponds to infinite distance in phase space, $\hat{U}^{2}+\hat{\Omega}^{2}
\rightarrow \infty$. In terms of $\hat{r},\hat{\theta}$ the dynamical system (\ref{qfh}), (\ref{dgh})
becomes
\begin{eqnarray}
\frac{d\hat{r}}{dx}&=&\frac{\hat{r}\!-\!1}
{2\hat{r}}\Big\{2\hat{\alpha}\beta(1\!-\!\hat{r})^{2}+2\hat{\alpha}(1\!-\!2\hat{r})
+\hat{r}^{2}\big[2(1\!+\!\hat{\alpha})\!+\!\cos{2\hat{\theta}}
\!-\!(\beta\!+\!1)\sin{2\hat{\theta}}\big]\Big\}\\
\label{eksf}
\frac{d\hat{\theta}}{dx}&=&\frac{1}{2\hat{r}^{2}}\Big[2\hat{\alpha}(1\!-\!\hat{r})^{2}(\tan{\hat{\theta}}\!-\!\beta
\cot{\hat{\theta}})+(\beta\!+\!1)\hat{r}^{2}\cos{2\hat{\theta}}+\hat{r}^{2}
(\sin2{\hat{\theta}}\!+\!\beta\!-\!1)\Big]\,.
\label{oefe}
\end{eqnarray}
As $\hat{r}\simeq 1^{-}$ the leading terms of equation (\ref{oefe}) for $\frac{d\hat{\theta}}{dx}$ are
\begin{equation}
\frac{d\hat{\theta}}{dx}\simeq\frac{1}{2}\big[(\beta\!+\!1)\cos{2\hat{\theta}}
\!+\!\sin{2\hat{\theta}}\!+\!\beta\!-\!1\big]\,,
\label{fqwf}
\end{equation}
while at linear order $\frac{d\hat{r}}{dx}=0$. The critical points $\hat{\theta}_{\ast}$ at infinity
are obtained by setting $\frac{d\hat{\theta}}{dx}= 0$ in (\ref{fqwf}) and solving for $\hat{\theta}$,
thus
\begin{equation}
(\beta\!+\!1)\cos{2\hat{\theta}_{\ast}}\!+\!\sin{2\hat{\theta}_{\ast}}\!+\!\beta\!-\!1=0\,.
\label{ellw}
\end{equation}
This equation has for any $\beta>0$ two roots for $\hat{\theta}_{\ast}$, one positive and one
negative. Therefore, there are always two fixed points at infinity, one in the upper quadrant
of the $(\hat{r},\hat{\theta})$ plane and the other in the lower quadrant, as are also depicted
in the uncompactified plots of Fig. 1, 2, 3. Since $\frac{d\hat{r}}{dx}|_{\hat{\theta}_{\ast}}=0$ we
cannot rely on the linearized analysis to examine the stability of these fixed points and numerical
examination is needed. Indeed, it can be seen numerically that for the parameters of Fig. 1, 2 a stable fixed point exists
for $\hat{\theta}_{\ast}>0$ (for which it is $\frac{d}{d\hat{\theta}}(\frac{d\hat{\theta}}{dx})<0$) and an
unstable fixed point exists for $\hat{\theta}_{\ast}<0$ (for which it is
$\frac{d}{d\hat{\theta}}(\frac{d\hat{\theta}}{dx})>0$), in agreement with the phase portraits of Fig. 1, 2.

\section{General solutions}

Equation (\ref{wdh}) can be approximated in the case that $U,\Omega>0$ are large (compared to
the gravity coupling $\kappa$), or more precisely the dimensionless quantities
$\hat{U},\hat{\Omega}>0$ are large. It is obvious from Fig. 1, 2 that large values of $U,\Omega$
correspond to large $R$, something that will be verified after the solution is found. More precisely,
for
\begin{equation}
\Big|\Omega-\frac{3}{2}U\Big|\gg\frac{|\alpha|}{U}\,\,\,\,\,\,\,,\,\,\,\,\,\,\,
\Big|U-\frac{1}{2\beta}\Omega\Big|\gg\frac{|\alpha|}{\Omega}
\label{uem}
\end{equation}
we get
\begin{equation}
\frac{d\Omega}{dU}=\frac{2\beta U-\Omega}{2\Omega-3U}\,,
\label{enz}
\end{equation}
which is a homogeneous equation. The above two inequalities are obviously consistent for
large $U,\Omega$, and moreover, these inequalities (together with (\ref{era})) will define the exact
$R$-domain where the solution
is valid. The puzzling situations with $\Omega\sim\frac{3}{2}U$ or $\Omega\sim 2\beta U$ ($U$ large)
do not occur since they provide for large $R$ that $\frac{d\Omega}{dU}\rightarrow \infty$ or 0
respectively, which is seen from Fig. 1, 2 not to be the case. The above inequalities set restrictions
in the allowed region of the 3-dimensional space $(x,U,\frac{dU}{dx})$ of the initial second order
differential equation (\ref{ujf}) where the orbits reside. Due to this, although we will find a general
large-distance (non-linearized) solution with the correct (maximum) number of integration constants,
the solution will fail to describe other regions in the space of initial data which also provide
large-distance solutions. This is obvious from the dynamical systems analysis of the previous section.

Assuming, for example, the parameter $|\hat{\alpha}|$ to be sufficiently small,
which means $|\Xi|=|\xi|\kappa^{2}\nu^{2} \gg 1$, the inequalities can more easily be satisfied and the
validity of the approximation becomes even more extended. At the same time the value of the parameter
$\beta$ is very close to $27/4$. Such values $|\Xi|\gg 1$ for the parameters were
argued in \cite{Kofinas:2015hla} that could in principle reduce a large value of the vacuum energy to a
small effective cosmological constant. Moreover, for such $\Xi$ the inequality (\ref{era}) becomes
$R\gg 2\sqrt{|\xi|}$, so this inequality is determined by the length scale defined by the non-minimal
coupling $\xi$. Additionally, since $\nu$ is related to the integration constant of the scalar field
$\phi$, it is expected to take large values for macroscopic solutions, thus even for small $|\xi|$ it
can be $|\Xi|\gg 1$. However, we do not restrict our analysis only in this range of parameters.

Setting
\begin{equation}
\Phi=\frac{\Omega}{U}>0\,,
\label{wdn}
\end{equation}
equation (\ref{enz}) becomes
\begin{equation}
U\frac{d\Phi}{dU}+2\frac{\Phi^{2}\!-\!\Phi\!-\!\beta}{2\Phi\!-\!3}=0\,,
\label{wdr}
\end{equation}
which is separable. Setting
\begin{equation}
\Psi=\Phi-\frac{1}{2}
\label{sdd}
\end{equation}
(with $\Psi>-\frac{1}{2}$), equation (\ref{wdr}) becomes
\begin{equation}
U\frac{d\Psi}{dU}+\frac{\Psi^{2}\!-\!\gamma^{2}}{\Psi\!-\!1}=0\,,
\label{hes}
\end{equation}
where $\gamma=\sqrt{\beta\!+\!\frac{1}{4}}>\frac{1}{2}$ (and thus $\Psi>-\gamma$).
Equation (\ref{hdh}) which makes the connection with the radial variable $x$ (or $R$)
takes the form
\begin{equation}
\frac{d\Psi}{dx}=\frac{\Psi^{2}\!-\!\gamma^{2}}{\Psi\!-\!1}
\Big(1\!-\!\Psi\!+\!\frac{\alpha}{U^{2}}\Big)\,.
\label{sqh}
\end{equation}
The parameter $\alpha$, which had temporarily disappeared in (\ref{enz}), came up here again.
The quantity $U^{2}$ in (\ref{sqh}) will be found as a function of $\Psi$ by integrating (\ref{hes}).
Indeed, the solution of (\ref{hes}) is
\begin{equation}
U^{2}=C\,
(\Psi\!+\!\gamma)^{-\frac{1}{\gamma}-1}\,|\Psi\!-\!\gamma|^{\frac{1}{\gamma}-1}\,,
\label{rhj}
\end{equation}
where $C>0$ is integration constant that distinguishes the solutions in the $(U,\Omega)$ space.
In terms of $\Omega$ the first integral (\ref{rhj}) takes the form
\begin{equation}
\Big{[}\Omega+\Big(\gamma-\frac{1}{2}\Big)U\Big{]}^{1+\frac{1}{\gamma}}\,
\Big{|}\Omega-\Big(\gamma+\frac{1}{2}\Big)U\Big{|}^{1-\frac{1}{\gamma}}=C\,.
\label{sdr}
\end{equation}
From (\ref{rhj}) it is seen that only for $\Psi\simeq \gamma>1$ it is $U\rightarrow \infty$, in
agreement with Fig. 1, 2. Therefore, we focus on $\gamma>1$ (which means $\beta>\frac{3}{4}$, while
$\alpha<\frac{\kappa^{2}}{2}$ or $\alpha>\kappa^{2}$) where $U$ can indeed extend to infinity.
Furthermore, since $\Omega=(\Psi+\frac{1}{2})U$, for such $\gamma$ there are solutions where both
$U,\Omega$ can extend to infinity. From (\ref{rhj}) it is seen that there are two classes of solutions,
one with $\Psi>\gamma$ and the other with $\Psi<\gamma$ (these solutions are represented
respectively by the left/right orbits of the upper quadrants in Fig. 1, 2). From (\ref{hes}) it is seen
that for the solutions with $\Psi>\gamma$ it is $\Psi$ a monotonically decreasing function of $U$.
To see if the above solutions correspond to large $R$, we set $\Psi\simeq \gamma$ in the integral
(\ref{rhj}) or even the equation (\ref{sqh}). Indeed, we get
$R\simeq R_{0}/|\Psi-\gamma|^{\frac{1}{2\gamma}}$, so the region $R\rightarrow \infty$ in included
and the metric can be extended to large distances. Obviously, $\Psi$ is not necessarily close to
the asymptotic value $\gamma$ in the range of applicability of the solution and the constraints
(\ref{ela}), (\ref{uem}) will determine the exact domain of $R$.

The metric (\ref{jsb}) becomes
\begin{equation}
ds^{2}=-N^{2}dt^{2}+\frac{|\Psi\!-\!\gamma|^{1-\frac{1}{\gamma}}\,
(\Psi\!+\!\gamma)^{1+\frac{1}{\gamma}}\,dR^{2}}{\nu^{2}CR^{2}}+R^{2}d\Omega_{2}^{2}\,,
\label{arg}
\end{equation}
where the relation of $R,\Psi$ in (\ref{arg}) is found from (\ref{sqh})
\begin{equation}
R(\Psi)=R_{0}\,e^{\int\frac{\Psi-1}{\Psi^{2}-\gamma^{2}}\frac{d\Psi}{1-\Psi+\alpha U^{-2}}}=
R_{0}\,e^{\int\frac{\Psi-1}{\Psi^{2}-\gamma^{2}}\frac{d\Psi}
{1-\Psi+\frac{\alpha}{C}(\Psi+\gamma)^{1+\gamma^{-1}}|\Psi-\gamma|^{1-\gamma^{-1}}}}\,,
\label{gck}
\end{equation}
with $R_{0}>0$ integration constant. Integration (\ref{gck}) cannot be performed analytically for
arbitrary $\gamma$. However, we will see that this is not necessary because within our approximation,
equation (\ref{gck}) becomes simplified. Metric (\ref{arg}) has an implicit form through integrating
and inverting (\ref{gck}) for $\Psi(R)$. However, we can obtain the metric (\ref{arg}) in a more
explicit form without the need for inversion (although not in the standard radial gauge) as follows
\begin{equation}
ds^{2}=-N^{2}dt^{2}+\frac{(\Psi\!-\!1)^{2}\,d\Psi^{2}}
{\nu^{2}C\,|\Psi\!-\!\gamma|^{1+\frac{1}{\gamma}}\,
(\Psi\!+\!\gamma)^{1-\frac{1}{\gamma}}\,
(1\!-\!\Psi\!+\!\alpha U^{-2})^{2}}+R(\Psi)^{2}d\Omega_{2}^{2}\,.
\label{shk}
\end{equation}
Since the master equation (\ref{jsi}) is of second order, there are two extra integration constants
$C,R_{0}$ in the solution (\ref{gck}), beyond the integration constant $\nu$. Therefore, after the
lapse function $N$ is found, the metric (\ref{arg}) or (\ref{shk}) will be the general approximate
solution in the domain of its validity.

The lapse metric function $N$ can be obtained from the equations (\ref{gij}), (\ref{dlnRN21}) from
where we obtain
\begin{equation}
\frac{d\ln{(RN^{2})}}{dx}=\frac{\epsilon\tilde{\eta}}{2\eta}\frac{1}{Y}
\Big(\frac{dY}{dx}\!+\!3Y\!+\!2\alpha\Big)\,,
\label{eif}
\end{equation}
with $\epsilon=\pm 1$ a sign symbol. Integration of (\ref{eif}) gives
\begin{equation}
N^{2}=\frac{\tilde{c}}{R}
\Big(R^{3}U^{2}e^{2\alpha\int\frac{dR}{RU^{2}}}\Big)^{\frac{\epsilon\tilde{\eta}}{2\eta}}\,,
\label{die}
\end{equation}
where $\tilde{c}>0$ is integration constant. Using (\ref{hdh}), (\ref{wdn}), (\ref{hes}) we find
\begin{equation}
N^{2}(R)=\frac{c}{R}\,\Big[\frac{R^{3}\,e^{J}}{|\Psi\!-\!\gamma|^{1-\frac{1}{\gamma}}
\,(\Psi\!+\!\gamma)^{1+\frac{1}{\gamma}}}\Big]^{\frac{\epsilon\tilde{\eta}}{2\eta}}\,,
\label{dke}
\end{equation}
where
\begin{equation}
J=2\alpha\int\frac{\Psi\!-\!1}{\Psi^{2}\!-\!\gamma^{2}}\,\frac{d\Psi}{\alpha\!+\!(1\!-\!\Psi)U^{2}}
=2\int\frac{\Psi\!-\!1}{\Psi^{2}\!-\!\gamma^{2}}\frac{d\Psi}{1+\frac{C}{\alpha}(1\!-\!\Psi)
|\Psi\!-\!\gamma|^{\frac{1}{\gamma}-1}(\Psi\!+\!\gamma)^{-\frac{1}{\gamma}-1}}
\label{uje}
\end{equation}
and $c=\tilde{c}\,C^{\frac{\epsilon\tilde{\eta}}{2\eta}}>0$ is a redefined integration constant.
The integration constant $c$ is not essential since it can be absorbed into a redefinition of the
time coordinate $t$.

The potential can be found from (\ref{wij}), after use of (\ref{hdh}), (\ref{wdn}), to be
\begin{eqnarray}
\eta V&\!=\!&-\Big(\Psi+\frac{1}{2}\Big)U^{2}+\frac{\alpha}{2}+\frac{1}{2\nu^{2}R^{2}}\\
&\!=\!& -\frac{C(\Psi\!+\!\frac{1}{2})}{|\Psi\!-\!\gamma|^{1-\frac{1}{\gamma}} \,(\Psi\!+\!\gamma)^{1+\frac{1}{\gamma}}}
+\frac{\alpha}{2}+\frac{1}{2\nu^{2}R^{2}}\,.
\label{heo}
\end{eqnarray}
Finally, the scalar field configuration can be determined from (\ref{dxdphiv}) using again (\ref{hdh}), (\ref{wdn})
\begin{eqnarray}
\phi&\!=\!&\phi_{1}+\epsilon_{1}\int\frac{\Psi\!-\!1}{\Psi^{2}\!-\!\gamma^{2}}\,
\frac{1}{1\!-\!\Psi\!+\!\alpha U^{-2}}\,\frac{d\Psi}{U}\\
&\!=\!&\phi_{1}+\frac{\epsilon_{1}}{\sqrt{C}}\int\frac{\Psi\!-\!1}{\Psi^{2}\!-\!\gamma^{2}}\,
\frac{|\Psi\!-\!\gamma|^{\frac{1}{2}(1-\frac{1}{\gamma})} \,(\Psi\!+\!\gamma)^{\frac{1}{2}(1+\frac{1}{\gamma})}}
{1\!-\!\Psi\!+\!\frac{\alpha}{C}|\Psi\!-\!\gamma|^{1-\frac{1}{\gamma}} \,(\Psi\!+\!\gamma)^{1+\frac{1}{\gamma}}}\,d\Psi\,,
\label{ewk}
\end{eqnarray}
where $\epsilon_{1}$ is another $\pm$ sign and $\phi_{1}$ is an integration constant.

In terms of $\Psi$ the inequalities (\ref{uem}) are written as
\begin{equation}
\frac{(\Psi\!+\!\gamma)^{1+\gamma^{-1}}\,|\Psi\!-\!\gamma|^{1-\gamma^{-1}}}{|\Psi-1|}\ll\frac{C}{|\alpha|}
\,\,\,\,\,\,,\,\,\,\,\,\,2\beta\frac{(\Psi\!+\!\gamma)^{1+\gamma^{-1}}\,|\Psi\!-\!\gamma|^{1-\gamma^{-1}}}
{(\Psi\!+\!\frac{1}{2})\,|\Psi\!-\!(2\gamma^{2}\!-\!1)|}\ll \frac{C}{|\alpha|}\,.
\label{eot}
\end{equation}
From the second condition of (\ref{eot}) it is seen that the branch with $\Psi>\gamma$ has a pole
at $\Psi=2\gamma^{2}-1>\gamma$. Therefore choosing a value for $\frac{C}{|\alpha|}$ both conditions
(\ref{eot}) will be satisfied for a solution which
is defined from $\Psi=\gamma$ (for $R\rightarrow \infty$) up to some value smaller than $2\gamma^{2}-1$
(that corresponds to the minimum $R$). If $\frac{C}{|\alpha|}$
is sufficiently large, $\Psi$ can approach the value $2\gamma^{2}-1$. Similarly, from the first condition
of (\ref{eot}) the branch with $\Psi<\gamma$ has a pole at $\Psi=1$. Therefore choosing a value for
$\frac{C}{|\alpha|}$ both conditions (\ref{eot}) will be satisfied for a solution which is defined
from some value larger than 1 (that corresponds to
the minimum $R$) up to $\Psi=\gamma$ (for $R\rightarrow \infty$), since from (\ref{hes}) $\Psi$ is now
an increasing function of $U$. If $\frac{C}{|\alpha|}$ is sufficiently large, $\Psi$ can approach the
value 1.

Hopefully, it is obvious that the first of the conditions in (\ref{eot}) assures that the quantity
$\frac{\alpha}{C}(\Psi+\gamma)^{1+\gamma^{-1}} |\Psi-\gamma|^{1-\gamma^{-1}}$ in the integral
(\ref{gck}) is negligible. Then, the integration can be done and it gives
\begin{equation}
R=R_{0}\Big|\frac{\Psi\!+\!\gamma}{\Psi\!-\!\gamma}\Big|^{\!\frac{1}{2\gamma}}\,.
\label{dff}
\end{equation}
Accordingly from equation (\ref{rhj})
\begin{equation}
U=\frac{\sqrt{C}}{2\gamma}\Big(\frac{R}{R_{0}}\Big)^{\!\gamma-1}
\,\Big[1\!-\!s\Big(\frac{R_{0}}{R}\Big)^{\!2\gamma}\Big]\,,
\label{eij}
\end{equation}
where $s=\text{sgn}(\Psi-\gamma)$ discerns the two branches with $\Psi>\gamma$ or $\Psi<\gamma$.
In the branch with $\Psi>\gamma$, it
is seen from (\ref{dff}) that as $\Psi$ increases, the radius $R$ decreases, while in the branch with
$\Psi<\gamma$, as $\Psi$ increases, $R$ also increases, in accordance with the previous analysis.

Finally, the conditions (\ref{eot}) take the following form in terms of $R$
\begin{eqnarray}
&&\Big(\frac{R}{R_{0}}\Big)^{2\gamma}\!-\!s\gg 2\gamma\frac{|\alpha|}{C}\Big(\frac{R}{R_{0}}\Big)^{\!2}
\,\,\Big|1\!-\!\frac{\gamma\!+\!1}{2\gamma}\Big[1\!-\!s\Big(\frac{R_{0}}{R}\Big)^{2\gamma}\Big]\Big|^{-1}
\label{efo}\\
&&\Big(\frac{R}{R_{0}}\Big)^{2\gamma-2}\gg 2\beta\frac{|\alpha|}{C}
\,\Big{\{}1\!-\!\frac{2\gamma\!-\!1}{4\gamma}\Big[1\!-\!s\Big(\frac{R_{0}}{R}\Big)^{2\gamma}\Big]\Big{\}}^{\!-1}
\,\,\Big|1\!-\!\frac{2\gamma\!+\!4\beta\!-\!1}{4\gamma}\Big[1\!-\!s\Big(\frac{R_{0}}{R}\Big)^{2\gamma}\Big]\Big|^{-1}
\label{qil}
\end{eqnarray}
respectively. For $s=1$, these inequalities say that $R$ can become as close to $R_{0}$ as we wish,
given that $\beta|\alpha|/C$ is sufficiently small. Since $\beta>\frac{3}{4}$ this last condition
means $|\alpha|/C\ll 1$ and the inequalities (\ref{uem}) are satisfied practically for any $R$ larger
than $R_{0}$. Furthermore, if $R_{0}\gg|\nu|^{-1}|\alpha|^{-1/2}$, the condition (\ref{era})
is satisfied for any $R$ outside $R_{0}$. Therefore, all the conditions have been satisfied.
For example, if $|\xi|\kappa^{2}\nu^{2}\gg 1$, the condition $|\alpha|/C\ll 1$ becomes
$4C|\xi|\nu^{2}\gg 1$ which is satisfied if $C$ is not particularly small; moreover, for
$R_{0}\gg 2\sqrt{|\xi|}$, the condition (\ref{era}) is satisfied.

To proceed with the solution, now the integral $J$ in (\ref{uje}) is found to be
$J\propto \frac{\alpha}{C}\int(\frac{\Psi+\gamma}{|\Psi-\gamma|})^{\frac{1}{\gamma}}d\Psi$, which
around the  dangerous for divergence point $\Psi=\gamma$ gives
$J\propto \frac{\alpha}{C} |\Psi-\gamma|^{1-\gamma^{-1}}$. Since $\gamma>1$, it is
$|\Psi-\gamma|^{1-\gamma^{-1}}$ finite and it will be $J\simeq 0$ for $|\alpha|/C\ll 1$.
Using (\ref{dff}), the metric (\ref{arg}) becomes
\begin{equation}
ds^{2}=-\Big(\frac{R}{R_{0}}\Big)^{\!\zeta\sqrt{\frac{3(2\gamma+1)}{2\gamma-1}}-1}
\left[1\!-\!s\Big(\frac{R_{0}}{R}\Big)^{\!2\gamma}\right]^{\zeta\sqrt{\frac{3}{\beta}}}d\hat{t}^{2}
+\frac{1}{\hat{C}R_{0}^{2}}\,
\frac{dR^{2}}{\big(\frac{R}{R_{0}}\big)^{\!2\gamma}
\left[1\!-\!s(\frac{R_{0}}{R})^{2\gamma}\right]^{2}}+R^{2}d\Omega_{2}^{2}\,,
\label{jdd1}
\end{equation}
where $\zeta=\epsilon\,\text{sgn}[(1\!-\!2\Xi)(1\!-\!6\Xi)]=\pm 1$,
$\hat{C}=\frac{C\nu^{2}}{4\gamma^{2}}>0$ and
$\hat{t}=\sqrt{c}\,(2\gamma)^{-\zeta\sqrt{\frac{3}{4\beta}}}\,R_{0}^{\frac{3}{2}\zeta\sqrt{\frac{3}{4\beta}}
-\frac{1}{2}} t$.

We will be interested in the solution with $s=1$ ($\Psi>\gamma$) which presents a pole.
As explained above this solution is valid for distances immediately outside $R_{0}$
and for this reason, $R_{0}$ for $\zeta=1$ will be called horizon. The metric (\ref{jdd1}) in this case
takes the form
\begin{equation}
ds^{2}=-\Big(\frac{R}{R_{0}}\Big)^{\!\sqrt{\frac{3(2\gamma+1)}{2\gamma-1}}-1}
\left[1\!-\!\Big(\frac{R_{0}}{R}\Big)^{\!2\gamma}\right]^{\sqrt{\frac{3}{\beta}}}d\hat{t}^{2}
+\frac{1}{\hat{C}R_{0}^{2}}\,
\frac{dR^{2}}{\big(\frac{R}{R_{0}}\big)^{\!2\gamma}
\left[1\!-\!(\frac{R_{0}}{R})^{2\gamma}\right]^{2}}+R^{2}d\Omega_{2}^{2}\,,
\label{sjf}
\end{equation}
where for convenience we remind that $\hat{C}>0$, $\gamma=\sqrt{\beta\!+\!\frac{1}{4}}>1$ and
$\beta=\frac{3}{4}\big(\frac{4\alpha}{\kappa^{2}}-3\big)^{2}>\frac{3}{4}$
with $\alpha<\frac{\kappa^{2}}{2}$ or $\alpha>\kappa^{2}$ (the parameter $\alpha$ is related to the
coupling $\xi$ and the scalar field integration constant $\nu$ through the combination
$\xi\kappa^{2}\nu^{2}$).
The analysis of the curvature invariants reveals that these are all finite at the horizon $R_{0}$
and diverge at infinity. The integration constant $R_{0}$ is expected to be related with the mass
of the black hole. Asymptotically the metric (\ref{sjf}) takes the form
\begin{equation}
ds_{\infty}^{2}=-\Big(\frac{R}{R_{0}}\Big)^{\!\sqrt{\frac{3(2\gamma+1)}{2\gamma-1}}-1}d\hat{t}^{2}
+\frac{1}{\hat{C}R_{0}^{2}}\,
\frac{dR^{2}}{\big(\frac{R}{R_{0}}\big)^{\!2\gamma}}+R^{2}d\Omega_{2}^{2}\,.
\label{sjv}
\end{equation}
This metric defines an asymptotic behaviour different than that of AdS space. Similarly to AdS, the
lapse function also goes to infinity for large distances, but here the scaling behavior is different.
More precisely, note that the exponent $2\gamma$ is larger than 2 and approaches the value 2 as
$\gamma$ approaches 1. On the contrary, the exponent $\sqrt{\frac{3(2\gamma+1)}{2\gamma-1}}-1$ is
smaller than 2 and positive, and approaches also the value 2 as $\gamma$ approaches 1. Therefore, for
$\gamma$ very close to 1 the asymptotic metric (\ref{sjv}) gets close to AdS, while as $\gamma$ departs
from the value 1, the metric gets a different structure.
For example, for $\gamma=7/2$, the lapse function in (\ref{sjv}) becomes proportional to $R$ (to give
an astrophysical perspective, the metric (\ref{sjv}) seems to provide extra attraction at large
distances, while fittings of linear potentials for exponential galactic disks have been
shown to explain the almost flat galactic rotation curves \cite{Mannheim:1992vj} and such potentials
yield galactic stability without the need of dark matter \cite{Christodoulou}). For a circular orbit at
constant radius $R$ the conditions $EN^{2}=1-\frac{R}{2N^{2}}\frac{dN^{2}}{dR}$,
$j^{2}=\frac{R^{3}}{2N^{4}} \frac{dN^{2}}{dR}$ have to be satisfied \cite{Weinberg}, where the constant
$E$ is the energy of the moving particle ($E>0$ for material particles, $E=0$ for photons) and the
constant $j$ is related to the angular momentum. For the metric (\ref{sjv}) both conditions are
satisfied since $\gamma>1$, therefore at large distances circular orbits are supported.

Making the transformation
\begin{equation}
\rho=R^{\gamma}\,,
\label{kwl}
\end{equation}
the metric (\ref{sjf}) takes the form
\begin{equation}
ds^{2}=\rho^{-\vartheta}\Big[-\rho^{2z}\Big(1\!-\!\frac{R_{0}^{2\gamma}}{\rho^{2}}\Big)
^{\!\!\sqrt{\frac{3}{\beta}}}d\hat{\tau}^{2}+\frac{R_{0}^{2(\gamma-1)}}{\hat{C}\gamma^{2}}
\frac{d\rho^{2}}{\rho^{2}\big(1\!-\!\frac{R_{0}^{2\gamma}}{\rho^{2}}\big)^{2}}
+\rho^{2}d\Omega_{2}^{2}\Big]\,,
\label{wik}
\end{equation}
where
\begin{equation}
0<\vartheta=2\Big(1\!-\!\frac{1}{\gamma}\Big)<2\,\,\,\,\,\,\,\,,\,\,\,\,\,\,\,\,
0<z=1-\frac{1}{2\gamma}\Bigg[3\!-\!\sqrt{\frac{3(2\gamma\!+\!1)}{2\gamma\!-\!1}}\,\Bigg]<1
\label{wfl}
\end{equation}
and
$\hat{\tau}=R_{0}^{\frac{1}{2}\big[1-\sqrt{\frac{3(2\gamma+1)}{2\gamma-1}}\big]}\,\hat{t}$.
The metric (\ref{wik}) is a hypescaling violating black hole \cite{Balasubramanian:2008dm} with
spherical horizon topology.
A hyperscaling violating black hole is a generalization of the Lifshitz black hole where $\vartheta=0$
(in our solution the metric can never asymptote a Lifshitz spacetime).
The Lifshitz metric arises as solution of gravity theories with negative cosmological
constant coupled to appropriate matter with the simplest such theory also including an abelian
gauge field \cite{Kachru:2008yh} (a pure Einstein gravity with cosmological constant
cannot produce an anisotropy in spacetime). Such metrics have also been found as solutions in string
theory and supergravities which arise from string constructions \cite{Hartnoll:2009ns}.
Effective gravity theories with a Maxwell as well as a dilaton field (in general a scalar field with
a non-trivial potential) are quite rich and have been shown to contain hyperscaling violating
solutions \cite{Gubser:2009qt}.
Note that the majority of the Lifshitz or hyperscaling violating Lifshitz
solutions and the corresponding black holes in the literature have planar (horizon) topology and are
assumed to have direct correspondence with condensed matter physics through the AdS/CFT conjecture
(for a spherical horizon topology to be obtained an extra gauge field should be added).
On the contrary, the spherical symmetry found here may offer to the solution some significance
at local astrophysical objects at large distance scales.
In general it is not easy for a given theory to possess Lifshitz
or hyperscaling violating solutions, and as referred, for example the introduction of some extra matter
source or higher order gravity theories are required. Here our scalar-torsion theory is an
additional case which provides such solutions and notably these solutions are general, while in most
of the known cases the arising such black holes are special solutions.
According to the standard notation,
$\vartheta$ is the hyperscaling violation exponent, while $z$ is the dynamical critical
exponent which indicates the anisotropy between time and space. For our solution the values that
these parameters can take are seen from the conditions (\ref{wfl}). The asymptotic form of
(\ref{wik}) is
\begin{equation}
ds_{\infty}^{2}=\rho^{-\vartheta}\Big[-\rho^{2z}
d\hat{\tau}^{2}+\frac{R_{0}^{2(\gamma-1)}}{\hat{C}\gamma^{2}}
\frac{d\rho^{2}}{\rho^{2}}+\rho^{2}d\Omega_{2}^{2}\Big]\,.
\label{wim}
\end{equation}
The scaling transformation
$\hat{\tau}\rightarrow \lambda^{z}\hat{\tau},\rho\rightarrow \lambda^{-1}\rho,x_{i}\rightarrow
\lambda x_{i}$ does not act as an isometry for the metric (\ref{wim}), so (\ref{wim}) is not
scale invariant, but it transforms conformally as $ds_{\infty}^{2}\rightarrow \lambda^{\vartheta}
ds_{\infty}^{2}$.

In the context of AdS/CFT, a non-vanishing $\vartheta$ indicates a
hyperscaling violation in the dual field theory. In the four-dimensional framework we are working,
theories with hyperscaling at finite temperature have an entropy density which scales with temperature
as $S\sim T^{\frac{2}{z}}$. For hyperscaling violation there is a modified relationship
$S\sim T^{\frac{2-\vartheta}{z}}$, indicating that the system lives in an effective dimension
$d_{\text{eff}}=2-\vartheta$. For the present solution it is $0<d_{\text{eff}}=\frac{2}{\gamma}<2$.

The potential accompanying the black hole (\ref{sjf}) is found from (\ref{heo}) to be
\begin{equation}
V=-\frac{\hat{C}}{2\alpha}\Big(\frac{R}{R_{0}}\Big)^{\!2(\gamma-1)}
\Bigg[1\!-\!\Big(\frac{R_{0}}{R}\Big)^{\!2\gamma}\Bigg]\,
\Bigg[2\gamma\!+\!1\!+\!(2\gamma\!-\!1)\Big(\frac{R_{0}}{R}\Big)^{\!2\gamma}\Bigg]
+\frac{\nu^{2}}{2}+\frac{1}{2\alpha R^{2}}\,.
\label{jel}
\end{equation}
It is seen that at the horizon the potential is finite, while at infinity it diverges as
$V\sim (R/R_{0})^{2(\gamma-1)}$.
Finally, the scalar field associated with the metric (\ref{sjf}) is found
{\footnote{From the Gauss' recursion formula $c(c+1)\,_{2}F_{1}(a,b;c;z)-c(c+1)\,_{2}F_{1}(a,b;c+1;z)-
abz\,_{2}F_{1}(a+1,b+1;c+2;z)=0$ (p. 1010 of \cite{grad}) and $_{2}F_{1}(a,b;a;z)=(1-z)^{-b}$ \cite{abra},
we get the relation
$(a+1)\,_{2}F_{1}(a,b;a+1;z)+bz\,_{2}F_{1}(a+1,b+1;a+2;z)=(a+1)(1-z)^{-b}$. Since
$\frac{d}{dz}\,_{2}F_{1}(a,b;c;z)=\frac{ab}{c}\,_{2}F_{1}(a+1,b+1;c+1;z)$ \cite{abra}, we get
$\frac{d}{dz}[z^{a}\,_{2}F_{1}(a,b;a+1;\mu z)]=az^{-(1-a)}(1-\mu z)^{-b}$. For (\ref{wjf}) it is also
used the transformation formula $_{2}F_{1}(a,a;a+1;\frac{z}{z-1})=(1-z)^{a}\,_{2}F_{1}(a,1;a+1;z)$ \cite{abra}.}}
to be
\begin{equation}
\phi=\phi_{1}-\frac{\epsilon_{1}|\nu|}{(\gamma\!-\!1)\sqrt{\hat{C}}}
\Big(\frac{R_{0}}{R}\Big)^{\!\gamma-1}\,_{2}F_{1}\Bigg(\frac{\gamma\!-\!1}{2\gamma},
1;\frac{3\gamma\!-\!1}{2\gamma};\Big(\frac{R_{0}}{R}\Big)^{\!2\gamma}\Bigg)\,.
\label{wjf}
\end{equation}
Since the parameter $\gamma$ depends on the integration constant $\nu$ of the scalar field, the scalar
field (\ref{wjf}) is a primary hair. The scalar field diverges at the horizon and is finite
at infinity. This behaviour at infinity is different than the behaviour found in \cite{Kofinas:2015hla}
for AdS asymptotics, where the scalar field evolves logarithmically with distance.
At infinite distance the scalar field behaves to dominant order as
$\phi-\phi_{1}\simeq
-\frac{\epsilon_{1}|\nu|}{(\gamma-1)\sqrt{\hat{C}}}(\frac{R_{0}}{R})^{\!\gamma-1}$.
Then, we can find the corresponding behaviour of the potential $V(\phi)$ for $\phi-\phi_{1}\simeq 0$ to be
\begin{equation}
V(\phi)\simeq -\frac{2\gamma\!+\!1}{2(\gamma\!-\!1)^{2}}\,\frac{\nu^{2}}{\alpha(\phi\!-\!\phi_{1})^{2}}\,.
\label{sjd}
\end{equation}
Thus, the potential $V(\phi)$ close to the origin $\phi-\phi_{1}=0$ is very steep and can be either
positive or negative depending on the value of $\alpha$. At the opposite limit of distances close to
the horizon, the potential gets an almost constant value $V\simeq \frac{\nu^{2}}{2}$.

Concerning the components of the energy-momentum tensor $\mathcal{T}^{\nu}_{\,\,\,\,\mu}$,
although the scalar field diverges at the horizon, it is obvious that its energy density
$\mathcal{T}^{t}_{\,\,\,\,t}$ and the pressures $\mathcal{T}^{R}_{\,\,\,\,R}$,
$\mathcal{T}^{\theta}_{\,\,\,\,\theta}$ are finite there. Asymptotically these components
of $\mathcal{T}^{\nu}_{\,\,\,\,\mu}$ diverge.

{\it{To summarize, the most significant solution found in this section is described by the metric forms
(\ref{sjf}), (\ref{wik}) with the asymptotic structures (\ref{sjv}), (\ref{wim}). The solution develops
an horizon $R_{0}$ and $R$ is practically defined outside $R_{0}$ when
$R_{0}\gg|\nu|^{-1}|\alpha|^{-1/2}$ and $\frac{|\alpha|\nu^{2}}{4\gamma^{2}\hat{C}}\ll 1$.
The solution is supported by the potential (\ref{jel}) with the scalar field profile (\ref{wjf}).}}

The Hawking temperature $T$ of the black hole is determined by the periodicity of the Euclidean metric
$ds_{E}^{2}=g_{\tau\tau}d\tau^{2}+g_{RR}dR^{2}+R^{2}d\Omega_{2}^{2}
=N^{2}d\tau^{2}+K^{-2}dR^{2}+R^{2}d\Omega_{2}^{2}$ obtained by the analytic continuation
$t=-i\tau$. Thus, $T$ is given by the standard formulae
$4\pi T=(dg_{\tau\tau}/dR)/\sqrt{g_{\tau\tau}g_{RR}}|_{R_{0}}=\sqrt{K^{2}/N^{2}}\,d(N^{2})/dR|_{R_{0}}=
\sqrt{N^{2}/K^{2}}\,d(K^{2})/dR|_{R_{0}}=\sqrt{[d(N^{2})/dR]\,[d(K^{2})/dR]}|_{R_{0}}$,
given that $K^{2}$ vanishes at the horizon $R_{0}$. Due to the higher order
pole at the horizon, the temperature vanishes in analogy to the extremal Reissner-Nordstrom solution.

From the astrophysical point of view, the motion of a freely falling photon in the static isotropic
gravitational field (\ref{sphermetric}) is described by the equation
$(\frac{dR}{d\varphi})^{2}=R^{4}K^{2}(\frac{1}{j^{2}N^{2}}-\frac{1}{R^{2}})$.
Since the field is isotropic, the orbit of the particle can be considered to be confined to the
equatorial plane $\theta=\frac{\pi}{2}$.
At the distance $R_{\ast}$ of closest approach to the center it is $\frac{dR}{d\varphi}=0$, thus
$j^{2}N(R_{\ast})^{2}=R_{\ast}^{2}$. The deflection of the orbit from the direction of initial
incidence at infinite distance is
$\Delta\varphi=2|\varphi(R_{\ast})-\varphi_{\infty}|-\pi$, where $\varphi_{\infty}$ indicates the
incident direction. The larger the quantity $(\frac{dR}{d\varphi})^{2}$ in the
previous differential equation, the larger the deflection angle is. It can be easily seen that at
large distances $(\frac{dR}{d\varphi})^{2}$ for the metric (\ref{sjf}) is bigger than that of
the Schwarzschild metric. Thus, there is for our solution an extra deflection of light compared to
the Newtonian deflection. The situation of increased deflection compared to that caused by the
luminous matter has been well observed in galaxies or clusters of galaxies. In general, the metric
$ds^{2}=-R^{2a}dt^{2}+\frac{dR^{2}}{BR^{2b}}+R^{2}d\Omega_{2}^{2}=
\rho^{-\vartheta}(-\rho^{2z}dt^{2}+\frac{1}{Bb^{2}}\frac{d\rho^{2}}{\rho^{2}}
+\rho^{2}d\Omega_{2}^{2})$, $\vartheta=2-\frac{2}{b}$, $z=1+\frac{a-1}{b}$ ($\rho=R^{b}$) has an
extra deflection of light at large distances if $a<1,b-a>0$ which mean $(2-\vartheta)(1-z)>0$,
$(2-\vartheta)(2-2z+\vartheta)>0$. These conditions are obviously satisfied for the solution
found here.

\section{Conclusions} \label{Conclusions}

In this work we have extended our previous analysis \cite{Kofinas:2015hla} on the quest of finding
spherically symmetric solutions of a scalar-torsion theory. More precisely, we treat the torsion not
as an independent field but the teleparallel condition is imposed as a constraint, therefore the
corresponding connection is assumed to have vanishing curvature. This last condition has been
implemented by adopting the Weitzenb\"{o}ck connection whose coordinate components are a function of
the vierbein, so our torsion is the torsion of the Weitzenb\"{o}ck
connection and the dynamical object of the theory is solely the vierbein. The theory consists of
the Einstein gravity in its teleparallel representation
supplemented by a minimally coupled scalar field with potential. The novel additional
term of the theory is a non-minimal derivative coupling of the scalar field with the torsion
scalar (which is a particular quadratic combination of the torsion that provides under variation
with respect to the vierbein the Einstein tensor). The equations of motion are second order
differential equations and the theory (action and field equations) is both diffeomorphism and local
Lorentz  invariant, while after adopting the Weitzenb\"{o}ck connection in order to
proceed, the local Lorentz invariance is abolished.

The master equation found in \cite{Kofinas:2015hla}, determining the four-dimensional spherically
symmetric solutions, has here been elaborated and appropriately approximated in order to find
large-distance solutions. The solutions found here are non-linearized, so they are valid at any
distance larger than some length scale defined by the parameters/integration constants of the
problem (the region of the parameters/integration constants is not fine-tuned or particularly
narrow). A dynamical systems analysis has been performed and elucidates in the space of dynamical
variables describing all the local solutions the regions where the large-distance solutions inhabit.
This offers an intuition for the generality of the approximation method applied in order to obtain
the general solutions.

Special solutions have been found, among which black holes also, when the integration constant
of the scalar field takes a particular value. Asymptotically, these solutions get an AdS form.
The corresponding scalar field is a secondary hair. There are branches where the asymptotic value
of the potential coincides with the effective cosmological constant of AdS; also there are other
branches where the cosmological constant coming from the potential is positive (therefore it is
different than the effective cosmological constant of AdS) and this is due to the growth of a
non trivial profile of the scalar field at infinity after interacting with torsion.

Probably the most interesting solutions found are branches of general (from the point of view of
the number of integration constants) spherically symmetric solutions.
These solutions are attracted by stable fixed points at the asymptotic region of the phase portrait
of the dynamical system. Among them, there is the
general black hole solution described by the metrics (\ref{sjf}), (\ref{wik}) with their
asymptotic forms (\ref{sjv}), (\ref{wim}). The form (\ref{wik}) clearly shows that the solution
is a hyperscaling violating black hole with positive hyperscaling violation exponent and dynamical
critical exponent. Note that the solution obtained here is a general solution of our theory, while
the majority of the Lifshitz or hyperscaling violating solutions of differing theories in the
literature are special. Note also that the topology of the horizon here is spherical, while most
of the existing Lifshitz or hyperscaling violating black holes have planar horizon topology. Due to the
spherical symmetry the solution found may have astrophysical significance, e.g. may possess extra
attraction at large distances. Actually, we have found for a freely falling photon in the static
isotropic gravitational field an increase of its deflection compared to the Newtonian deflection.
The scalar field $\phi$ accompanying the solution is a primary hair which diverges at the horizon
and is finite at infinity. Notice, however, that in all the black hole solutions found in this work,
although the scalar field diverges at the location of the horizon, its energy density
and the pressures are finite there.

\[ \]
{{\bf Acknowlegements}}
I wish to thank Mokhtar Hassa\"{\i}ne, Elias Kiritsis and Miok Park for useful comments and discussions.


\end{document}